\documentclass{article}
\usepackage{frascatiphys,here,graphicx,subfigure}
\begin{document}
\title{Baryon spectroscopy in constituent quark models.}
\author{
J. Vijande        \\
{\em Departamento de F\'{\i}sica Te\'orica e IFIC. Universidad de Valencia - CSIC, Spain} \\
A. Valcarce        \\
{\em Departamento de F\'{\i}sica Fundamental, Universidad de Salamanca, Spain} \\
P. Gonzalez        \\
{\em Departamento de F\'{\i}sica Te\'orica e IFIC. Universidad de Valencia - CSIC, Spain} \\
H. Garcilazo        \\
{\em Escuela Superior de F\'{\i}sica y Matem\'aticas, Instituto Polit\'ecnico Nacional, Mexico.} \\
}
\maketitle
\baselineskip=11.6pt
\begin{abstract}
We present a study of the baryon spectra for all flavor sectors within a 
constituent quark model. We address some of the outstanding problems in 
baryon spectroscopy, as for example the spin splitting evolution for te 
different flavor sectors, the flavor independence of confinement and the 
missing state problem.

\end{abstract}

\baselineskip=14pt

\section{The light sector}

The complexity of Quantum Chromodynamics (QCD), the quantum field theory of the
strong interaction, has prevented so far a rigorous deduction of its
predictions even for the simplest hadronic systems. In the meantime while
lattice QCD starts providing reliable results, 
QCD-inspired models are useful tools to get some insight into many of the phenomena
of the hadronic world. One of the central issues to be
addressed is a quantitative description of the low-energy phenomena,
from the baryon-baryon interaction to the baryon spectra, still
one of the major challenges in hadronic physics.

Nowadays, we have at our disposal realistic quark models accounting
for most part of the one- and two-body low-energy hadron phenomenology.
Among the quark models found in the
literature\cite{Val05}, the ambitious project of a simultaneous description 
of the baryon-baryon interaction and the hadron
spectra in all the flavor sectors has only been undertaken by the constituent quark 
model of Ref.\cite{Vij04b}. The success in describing the 
properties of the strange and non-strange one and two-hadron
systems encourages its use as a guideline in order to assign parity 
and spin quantum numbers to already determined baryon
states as well as to predict still non-observed resonances. 

The results we are going to present have been obtained by solving
exactly the Schr\"odinger equation by the Faddeev method in momentum
space. The results are of similar quality to others present 
in the literature based on models specifically designed for the
study of the baryon spectra\cite{Val05b}. 
In the constituent quark model used in
this work the hyperfine splitting is shared between
pseudoscalar forces and perturbative QCD contributions, 
provided by the one-gluon exchange. 
In Table \ref{t2} we give the contribution of different pieces of the
interacting hamiltonian to the energy of several octet and decuplet baryons.
One observes that the hyperfine splittings are
controlled by the one-gluon exchange (OGE) and one-pion exchange (OPE) [one-kaon exchange (OKE)] 
potentials in the non-strange [strange] sector.
The OGE and OPE generate almost the experimental hyperfine splitting, the one-eta (OEE) and
one-sigma exchange (OSE) given a final small tune.
The expectation value of the OPE flavor operator for two light quarks 
is replaced by the similar effect of the OKE when a light and a strange quarks are involved.
They enhance in a similar way the hyperfine splitting
produced by the OGE.
The important effect of the OGE is observed when Table \ref{t2} is compared 
to Table II of Ref.\cite{Fur02}. The contribution of the pseudoscalar forces is much smaller
in our case, generating decuplet-octet mass differences of the order of
100$-$200 MeV, the remaining mass difference given by the OGE.
\begin{table}[tb]
\caption{Eigenvalue, in MeV, of the kinetic energy combined with different
contributions of the interacting potential. The subindexes in the 
potential stand for: $1=CON$, $2=1+OGE$, $3=1+OPE$,
$4=2+OPE$, $5=3+OKE$, $6=5+OEE$,
$7=6+OSE$. Experimental date is taken from the PDG.}
\label{t2}
\begin{center}
\begin{tabular}{c|ccccccc|c}
State & $V_1$ & $V_2$ & $V_3$ & $V_4$ & $V_5$ & $V_6$ & $V_7$ & Exp.\\ 
\hline
$N(1/2^+)$          &  1534  &  1254  &  1407  &  969  &   969  & 1030  &  939 & 939 \\  
$\Delta(3/2^+)$     &  1534  &  1314  &  1510  & 1291  &  1291  & 1283  & 1232 & 1232 \\  
$N^*(1/2^+)$        &  1787  &  1601  &  1716  & 1448  &  1448  & 1479  & 1435 & 1420--1470\\  
$N(1/2^-)$          &  1722  &  1530  &  1675  & 1422  &  1422  & 1447  & 1411 & 1515--1525\\  
$\Sigma(1/2^+)$     &  1679  &  1417  &  1674  & 1408  &  1326  & 1229  & 1213 & 1192.642$\pm$0.024\\  
$\Sigma(3/2^+)$     &  1679  &  1462  &  1673  & 1454  &  1437  & 1438  & 1382 & 1383.7$\pm$1.0\\  
$\Sigma^*(1/2^+)$   &  1983  &  1757  &  1931  & 1752  &  1703  & 1688  & 1644 & 1630--1690\\  
$\Sigma(1/2^-)$     &  1859  &  1677  &  1854  & 1671  &  1645  & 1634  & 1598 & $\approx 1620$\\  
$\Lambda(1/2^+)$    &  1679  &  1405  &  1600  & 1225  &  1171  & 1217  & 1122 & 1115.683$\pm$0.006\\  
$\Xi(1/2^+)$        &  1819  &  1557  &  1819  & 1557  &  1472  & 1446  & 1351 & 1321.31$\pm$0.13\\
$\Omega(3/2^+)$     &  1955  &  1743  &  1955  & 1743  &  1743  & 1728  & 1650 & 1672.45$\pm$0.29\\  
\end{tabular}
\end{center}
\end{table}

\section{The missing state problem}

Constituent quark models of baryon structure are based on the assumption of
effective quark degrees of freedom so that a baryon is a three-quark
color-singlet state. Lattice QCD in the quenched approximation shows 
out a $qq$ confining potential linearly rising
with the interquark distance\cite{Bal01}. This potential produces an
infinite discrete hadron spectrum. The implementation of this confining
force with OGE and/or Goldstone boson exchanges derived from chiral symmetry
breaking, or other effective interactions, turns out to be fruitful in the
construction of quark potential models providing a precise
description of baryon spectroscopy. However an outstanding problem remains 
unsolved: all models predict a proliferation of
baryon states at excitation energies above 1 GeV which are not
experimentally observed as resonances. This difference between the quark
model prediction and the data about the number of physical resonances is
known as the missing resonance problem.

Unquenched lattice QCD points out a string breaking in the static potential between 
two quarks\cite{Bal01} what should be properly incorporated in the phenomenological description of the 
high energy hadronic spectrum. The spontaneous creation of a quark-antiquark pair at the
breaking point may give rise to a breakup of the color flux tube between two quarks 
in such a way that the quark-quark potential does not rise with the interquark distance 
but it reaches a maximum saturation value. The simplest quark-quark screened potential, 
containing confinement and one-gluon exchange terms, reads:
\begin{equation}
V(r_{ij})=\frac{1}{2}\left[ \overline{\sigma }r_{ij}-\frac{\overline{\kappa }%
}{r_{ij}}+\frac{\hbar ^{2}\overline{\kappa _{\sigma }}}{m_{i}m_{j}c^{2}}%
\frac{e^{-r_{ij}/\overline{r_{0}}}}{\overline{r_{0}}^{2}r_{ij}}(\vec{\sigma
_{i}}\cdot \vec{\sigma _{j}})\right] \left( \frac{1-e^{-\mu \,\,r_{ij}}}{\mu
\,\,r_{ij}}\right) +\frac{\overline{M_{0}}}{3}  \label{eq1}
\end{equation}
where $r_{ij}$ is the interquark distance, $m_{i,j}$ the masses of the
constituent quarks, $\vec{\sigma}_{i,j}$ the spin Pauli operators, and $\overline{M_{0}}$ 
is a constant. The screening multiplicative factor appears between parenthesis on the right
hand side. $\mu $, the screening parameter, is the inverse of the saturation
distance and its effective value is fitted together with the other
parameters, $\overline{\sigma }$, $\overline{\kappa }$, and $\overline{%
\kappa _{\sigma }}$, to the spectrum.

For nonstrange baryons the model predicts quite approximately the number 
and ordering of the experimental states up to a mass of $2.3$ GeV\cite{Vij04,Gon03}. 

\begin{figure}[p]
    \begin{center}
        {\includegraphics[scale=0.53]{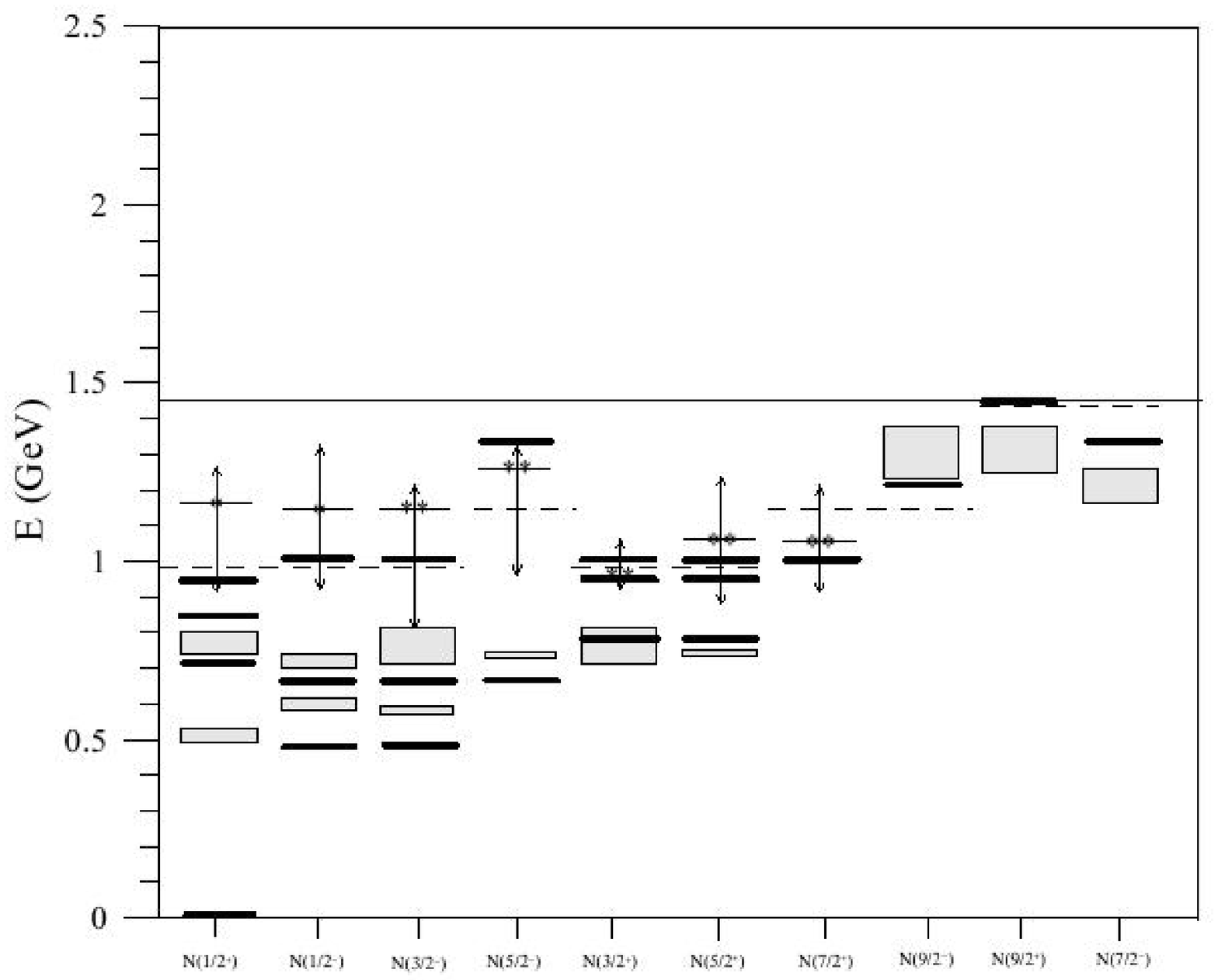}}
        {\includegraphics[scale=0.48]{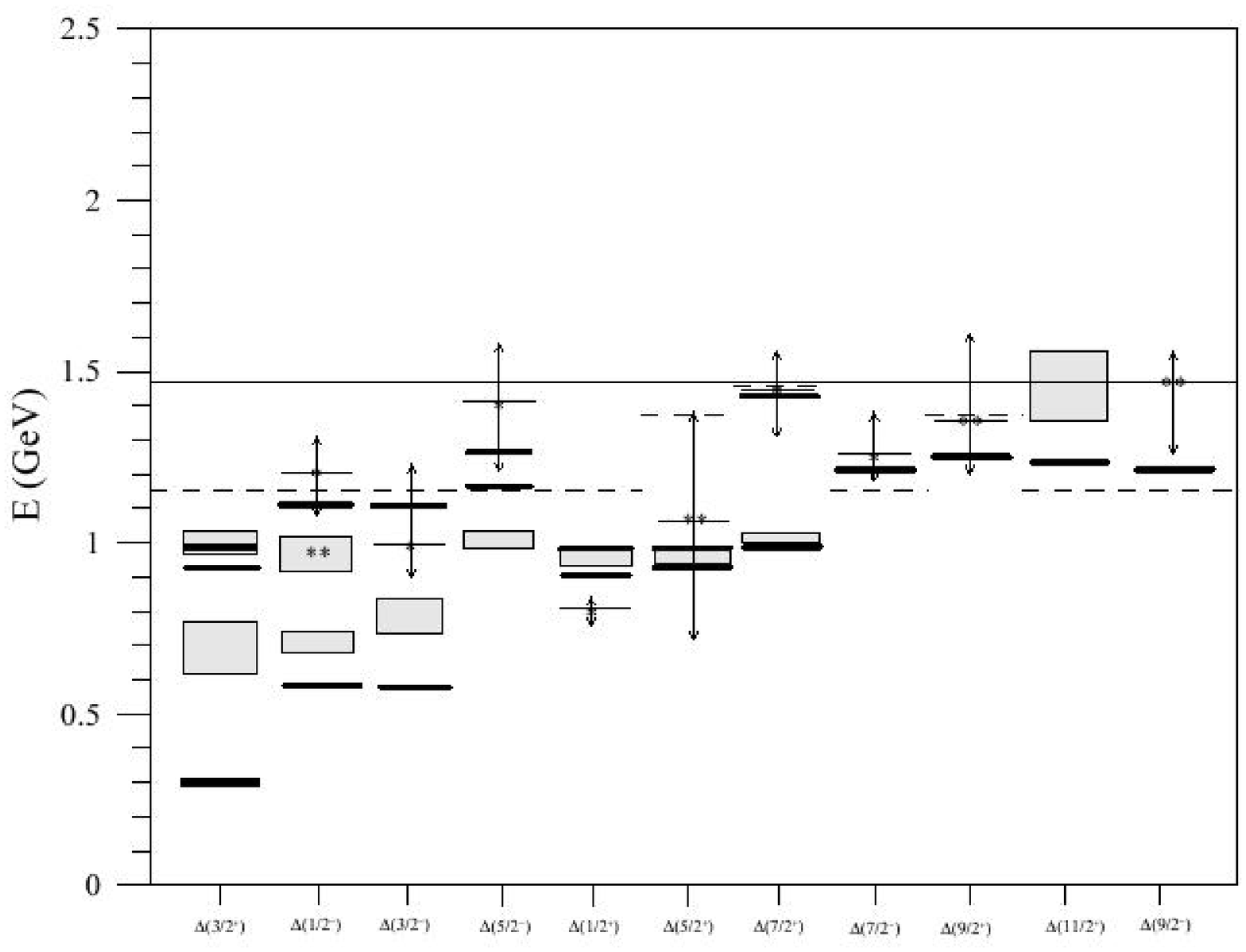}}
\caption{Relative energy nucleon (upper part) and $\Delta$ (lower part) spectra for the screened 
potential of Eq. (\protect\ref{eq2}) with the parameters of Ref.\protect\cite{Vij04}. The
thick solid lines represent our results. The shaded region, whose size
stands for the experimental uncertainty, represents the experimental data
for those states cataloged as $(\ast \ast \ast )$ or $(\ast \ast \ast \ast )$
states in the Particle Data Book. Experimental data
cataloged as $(\ast )$ or $(\ast \ast )$ states are shown by short thin
solid lines with stars over them and by vertical lines with arrows standing
for the experimental uncertainties. Finally, we show by a dashed line the $1q $ 
ionization threshold and by a long thin solid line the total threshold.}
\label{f1}
    \end{center}
\end{figure}

More recent lattice calculations\cite{Bal01} show that the 
$Q\overline{Q}$ potential saturates sharply for a breaking distance of 
the order of 1.25 fm corresponding to a saturation energy of about twice the $B$ meson 
($Q\overline{q})$ mass, indicating that the formation of two heavy-light
subsystems is energetically favored. A saturated quark-quark potential incorporating 
this effect can be parametrized as:

\begin{equation}
V(r_{ij})=\left\{\matrix{V_{sr}(r_{ij}) & & r_{ij} < r_{sat} \cr \sigma
r_{sat} & & r_{ij} \geq r_{sat}}\right. \, ,  \label{eq2}
\end{equation}
where 
\begin{equation}
V_{sr}(r_{ij})=\frac{1}{2}\left[ \sigma r_{ij}- \frac{\kappa }{r_{ij}}+\frac{%
\hbar ^{2}\kappa _{\sigma }} { m_{i}m_{j}c^{2}}\frac{e^{-r_{ij}/r_{0}}}{%
r_{0} ^{2}r_{ij}}(\vec{\sigma _{i}}\cdot \vec{\sigma _{j}})\right] + \frac{%
M_0}{3}
\end{equation}

\noindent whose parameters are given in Ref.\cite{Gon06}. The
calculation of the spectrum proceeds exactly in the same manner as in Ref.\cite{Vij04}, 
to which we refer for technical details. It is worth to remark 
that the presence, in the three-body problem, of
two-body thresholds (for only one quark to be released), 
apart from the absolute three-body ones (saturation energy),
may represent further constraints in the applicability limit of the model to
any particular channel. The results obtained are represented in Fig. \ref{f1}. 
As in Ref.\cite{Vij04} we have also included the
predicted states close above the thresholds.

The quality of the description of the light baryon spectra is remarkable 
since apart from keeping the same level of quality than in the low and 
medium-lying spectrum a perfect one to one correspondence between our
predicted states and the experimental resonances for any $J^{P}$ is obtained. 
Similar results are obtained using the screeend potential given in Eq.\ref{eq1}. 
The number and ordering of states remains unaltered.
The sharp potential tends quite generally to push upward the
highest energy states. In other words the screened 
potential is quite similar to the closest physical approach to a nonscreened 
potential, represented by the sharp interaction, that takes
effectively into account the effect of the baryon decay to open channels in
order to select the observed resonances.

\section{The heavy sector}

Since the discovery at BNL\cite{Caz75} and posterior confirmation 
at Fermilab\cite{Bal79} of the existence of charmed baryons in the late 70's, 
an increasing interest on heavy baryon spectroscopy arose. It became evident that baryons
containing heavy flavors $c$ or $b$ could play an important role in our
understanding of QCD. Since then, several new hadrons containing a single 
charm or bottom quark have been identified\cite{Eid04}. While the mass of these 
particles is usually measured 
as part of the discovery process, other quantum numbers such as the spin or
parity have often proved to be more elusive. For heavy baryons,
no spin or parity quantum numbers of a given state have been measured directly. 
Therefore, a powerful guideline for assigning quantum numbers 
to new states or even to indicate new states to look for is 
required by experiment. 

\begin{figure}[p]
    \begin{center}
        {\includegraphics[scale=0.45]{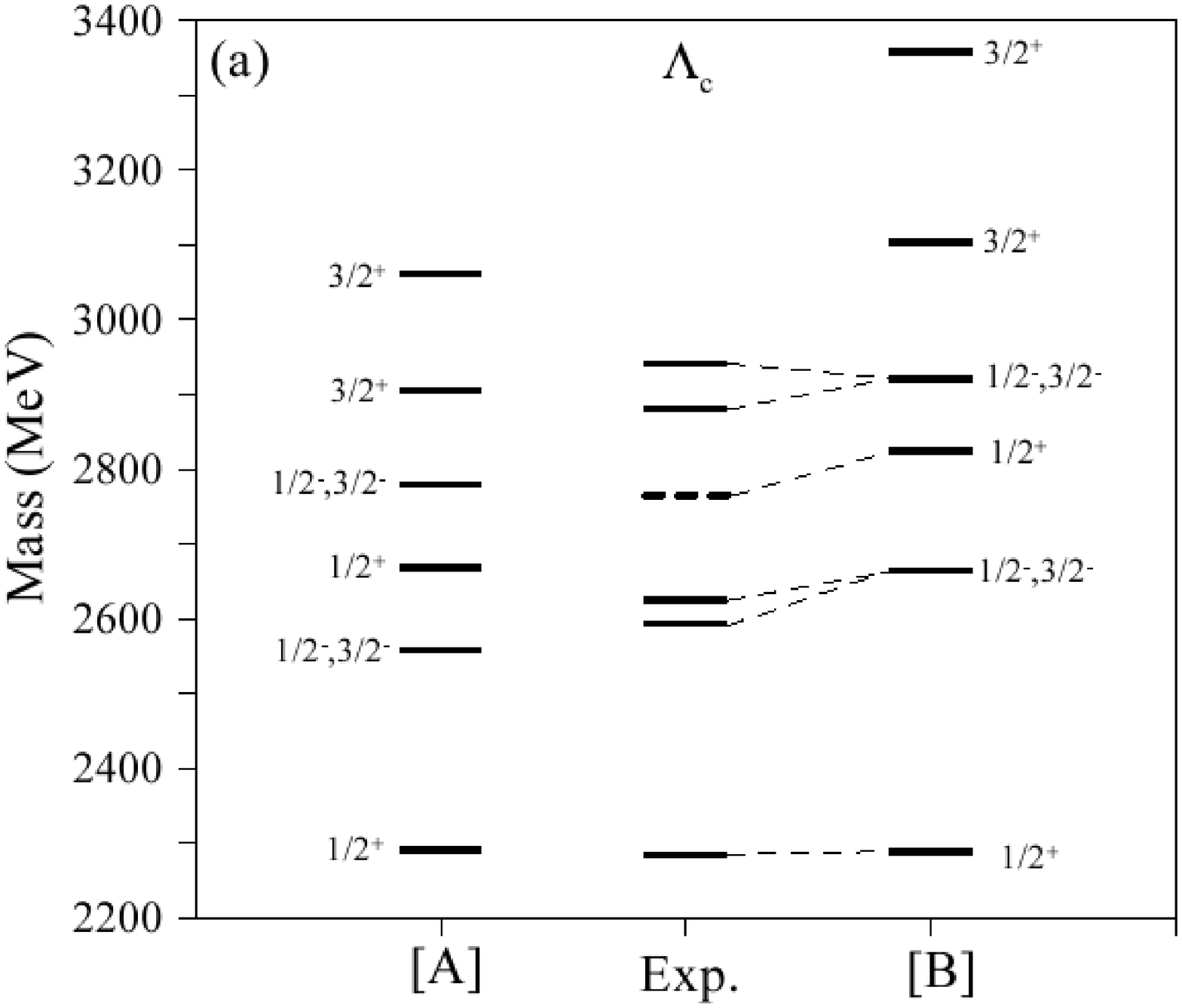}}
        {\includegraphics[scale=0.45]{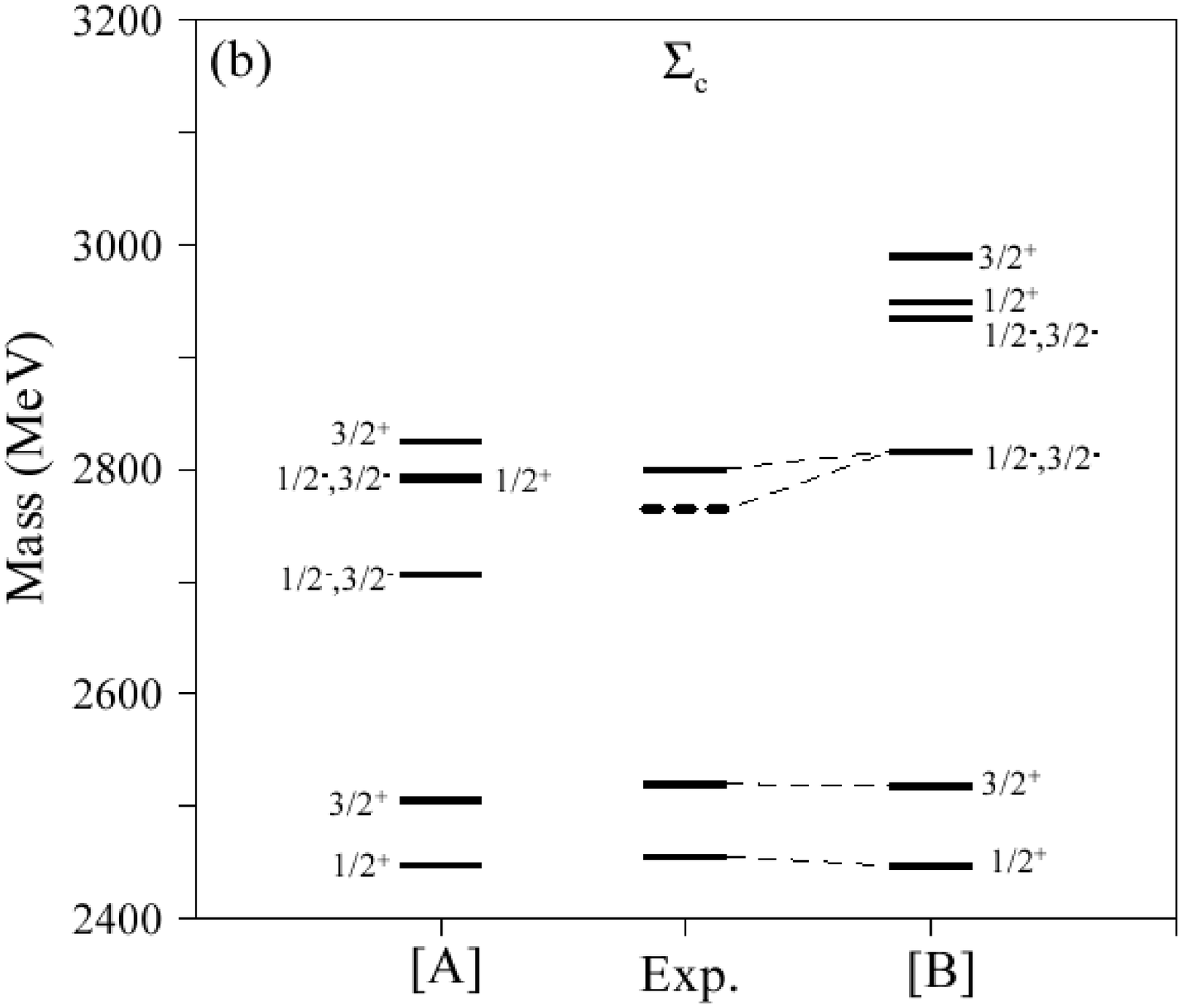}}
\caption{(a) Spectra of $\Lambda_c$ for two different confinement
strengths compared to experiment. (b) Same as (a) for $\Sigma_c$ states.}
\label{fig1}
    \end{center}
\end{figure}

Several criteria can be 
chosen to fit the confinement strength in the baryon spectra,
being the most usual ones to fit the energy splitting between the nucleon and its first radial
excitation (roper resonance) or to fit the splitting with its lowest orbital excitation (negative 
parity). We show the differences using both criteria in Fig. \ref{fig1}. On the left hand side we show
the spectra obtained in the first case, named [A], and on the right hand side the results obtained for the later, named [B].
A better agreement is observed with the model reproducing the orbital excitations of the
light baryon sector\cite{Vij07}. There is no experimental
state that we do not predict and there is no low-lying
theoretical resonance that has not been observed.
The recently discovered $\Sigma_c(2800)$\cite{Miz05} would correspond to an
orbital excitation with $J^P=1/2^-$ or $3/2^-$, any other correspondence being
definitively excluded. For $\Lambda_c$ baryons, the recently
confirmed as a $\Lambda_c$ state, $\Lambda_c(2880)$\cite{Aub06},
and the new state $\Lambda_c(2940)$\cite{Aub06} may constitute
the second orbital excitation of the $\Lambda_c$ baryon. Finally,
there is an state with a mass of 2765 MeV reported in Ref.\cite{Art01}
as a possible $\Lambda_c$ or $\Sigma_c$ state and also observed
in Ref.\cite{Miz05}. While the first reference (and also the PDG)
are not able to decide between a $\Lambda_c$ or a $\Sigma_c$
state, the second one prefers a $\Lambda_c$ assignment.
As seen in Fig. \ref{fig1}, this state may constitute the second member of the
first orbital excitation of $\Sigma_c$ states or the first radial
excitation of $\Lambda_c$ baryons. An experimental effort to
confirm the existence of this state and its decay modes would
help on the symbiotic process between experiment and theory
to disentangle the details of the structure of heavy baryons.


\begin{thebibliography}{99}
\bibitem{Val05} A. Valcarce, {\it et al},
	Rep. Prog. Phys. {\bf 68}, 965 (2005).

\bibitem{Vij04b} J. Vijande, F. Fern\'andez, and A. Valcarce,
	J. Phys. G {\bf 31}, 481 (2005).

\bibitem{Val05b} A. Valcarce, H. Garcilazo, J. Vijande,
		Phys. Rev. C{\bf 72}, 025206 (2005).

\bibitem{Fur02} M. Furuichi and K. Shimizu,
                Phys. Rev. C {\bf 65}, 025201 (2002).

\bibitem{Bal01} SESAM Coll., G.S. Bali, {\it et al}, 
		Phys. Rev. D {\bf 71}, 114513 (2005).

\bibitem{Vij04} J. Vijande, {\it et al},
		Phys. Rev. D {\bf 69}, 074019 (2004).

\bibitem{Gon03} P. Gonz\'{a}lez, {\it et al},
		Phys. Rev. D {\bf 68}, 034007 (2003).

\bibitem{Gon06} P. Gonz\'{a}lez, {\it et al},
		Eur. Phys. J. A{\bf 29}, 235 (2006).

\bibitem{Caz75} E.G. Cazzoli {\it et al.},
	Phys. Rev. Lett. {\bf 34}, 1125 (1975).

\bibitem{Bal79} C. Baltay {\it et al.},
	Phys. Rev. Lett. {\bf 42}, 1721 (1979).

\bibitem{Eid04} W.-M. Yao {\it et al.},
	 J. Phys. G {\bf 33}, 1 (2006).

\bibitem{Vij07} H. Garcilazo, J. Vijande, A. Valcarce,
	J. Phys. G{\bf 34}, 961 (2007). 

\bibitem{Miz05} Belle Coll., R. Mizuk {\it et al.},
	Phys. Rev. Lett. {\bf 94}, 122002 (2005).

\bibitem{Aub06} BABAR Coll., B. Aubert {\it et al.},
	Phys. Rev. Lett. {\bf 98}, 122011 (2007).

\bibitem{Art01} CLEO Coll., M. Artuso {\it et al.},
	Phys. Rev. Lett. {\bf 86}, 4479 (2001).

\end{thebibliography}
\end{document}